\documentclass{article}
\usepackage{amsfonts}

\usepackage{graphicx}
\usepackage{amsmath}

\input{tcilatex}

\begin{document}

\title{A tomographic setting for quasi-distribution functions}
\author{V.I. Man'ko$^a$, G. Marmo$^b$, A. Simoni$^b$, E.C.G. Sudarshan$^c$, \\
%EndAName
F. Ventriglia$^b$\\
{\footnotesize \textit{$^a$P.N.Lebedev Physical Institute, Leninskii
Prospect 53, Moscow 119991, Russia }}\\
{\footnotesize {(e-mail: \texttt{manko@na.infn.it})}}\\
\textsl{{\footnotesize {$^b$Dipartimento di Scienze Fisiche dell'
Universit\`{a} ``Federico II" e Sezione INFN di Napoli,}}}\\
\textsl{{\footnotesize {Complesso Universitario di Monte S. Angelo, via
Cintia, 80126 Naples, Italy}}}\\
{\footnotesize {(e-mail: \texttt{marmo@na.infn.it, simoni@na.infn.it,
ventriglia@na.infn.it})}}\\
{\footnotesize \textit{$^c$Department of Physics, University of Texas,
Austin, Texas 78712, USA}}\\
{\footnotesize {(e-mail: \texttt{sudarshan@physics.utexas.edu})}}}
\maketitle

\begin{abstract}
The method of constructing the tomographic probability distributions
describing quantum states in parallel with density operators is presented.
Known examples of Husimi-Kano quasi-distribution and photon number
tomography are reconsidered in the new setting. New tomographic schemes
based on coherent states and nonlinear coherent states of deformed
oscillators, including $q-$oscillators, are suggested. The associated
identity decompositions providing Gram-Schmidt operators are explicitly
given.

\end{abstract}

%\author[rus]{V.I. Man'ko\corauthref{cor1}},
%\ead{manko@na.infn.it}
%\author[na]{G. Marmo}, \ead{marmo@na.infn.it}
%\author[na]{A. Simoni\corauthref{cor1}},
%\corauth[cor1]{Corresponding authors} \ead{simoni@na.infn.it}
%\author[george]{E.C.G. Sudarshan}, \ead{sudarshan@physics.utexas.edu}
%\author[na]{F. Ventriglia}\ead{ventriglia@na.infn.it}
%\address[rus]{ P.N.Lebedev Physical Institute, Leninskii Prospect 53, Moscow 119991, Russia}
%\address[na]{Dip. Scienze Fisiche dell'Universit\`{a} Federico II e Sezione INFN di Napoli, \\
%Complesso Universitario Monte S. Angelo, I-80126 Naples, Italy}
%\address[george]{Department of Physics, University of Texas, Austin, Texas 78712, USA}

\section{Introduction}

There are several representations of quantum states providing the
possibility to present equivalent, but different in their form, formulations
of quantum mechanics \cite{AmJPhys}. The representations of quantum states
are based on different integral transforms of the density operator \cite
{Landau27,vonNeum27} taken in the position representation. The density
operator in the position representation is mapped by means of the integral
transforms either to Wigner quasi-distribution function \cite{Wig32}, or
Husimi-Kano $K-$function \cite{Hus40,Kano56}. In this paper we have decided
to keep up with the original notations of the pioneer papers on the subject.
Quasi-distributions are usually referred to as phase space representations
of quantum states. Another important phase space representation is related
to the Sudarshan's diagonal coherent state representation \cite{Sud63,Glau63}%
. Recently the tomographic representations of quantum states was suggested 
\cite{Ber-Ber,Vog-Ris,Mancini95} using the Radon integral transform of their
Wigner functions. The tomographic representation exhibits some specific
property in comparison with the other phase space representations. The
tomographic probability distributions (tomograms) associated with quantum
states are standard positive probability distributions. The mathematical
mechanism of constructing the tomographic probabilities in abstract Hilbert
spaces was elucidated in \cite{PLA2} for the finite dimensional case. Also
infinite dimensional Hilbert spaces were considered in the context of
mapping vectors and density operators onto tomographic probabilities in \cite
{PalPalme}.

The details of those tomographic approach constructions inspire to study
other possibilities to know better and create new schemes of description of
quantum states by positive functions. First of all, the Husimi-Kano $K-$%
function interpreted as quasi-distribution in the phase space representation
is a real non-negative bounded function of two variables, $q$ and $p$.
Strictly speaking, these variables cannot be interpreted as the actual
position and momentum since the uncertainty relations prohibit the existence
of a distribution in the phase space depending on variables which cannot be
measured simultaneously. Nevertheless, being a real non-negative and
normalized function on ``phase space'', it has a chance to be interpreted
from the point of view of the tomographic approach. One of the aims of the
present work is to extend our analysis of the mathematical mechanism of
constructing tomographic sets to extra situations, by including into the
construction properties of the coherent states \cite{SudKlaubook} and
properties of $K-$function and Sudarshan's diagonal coherent state
representation \cite{MehtaSud65} in the sense of their relation to
distributions (i.e., generalized functions) . We call this extension \textit{%
coherent state tomography}. From this point of view the Husimi-Kano $K-$%
function will be interpreted as a tomogram of a quantum state.

The coherent states are closely connected with linear vibrations (linear
harmonic oscillator). The deformed oscillators, e.g. $q-$ oscillators \cite
{Bied,Mc} and $f-$ oscillators \cite{SudPhysScr,Solim} are related at a
classical level with specific non-linear vibrations, so that non-linear
coherent states were introduced \cite{MVogel,Sudf} to describe the
corresponding states of a non-linear quantum oscillator, which yields in the
linearity limit the standard coherent state counterpart. Another goal of
this work is to develop the $f-$deformed version of the tomographic approach
and in particular to present, for a specific choice of the non-linearity
coded by the function $f=f_{q}$, the $q-$deformed coherent state tomography.

Also, by using $f-$deformed coherent states, we develop the $f-$deformed
version of the photon number tomography \cite{BanWog,WogW,TombesEPL} as well
as we obtain the identity decomposition for the deformed tomographies. In
fact, we are addressing the general problem of constructing a map from a
Hilbert space (or projective Hilbert space) onto a family of probability
distributions (tomograms).

As in the usual formulation of quantum mechanics there are several schemes
like Schr\"{o}dinger picture, Heisenberg picture, Dirac picture, even in the
tomographic approach exist several different schemes, like symplectic
tomography, photon number tomography, optical tomography and so on. Our last
aim is to point out the common general mechanism of constructing all these
tomographic schemes and in this way to extend the list of the tomographies
by including the deformed ones.

The paper is organized as follows. In section 2 a short account of our
construction of a tomographic setting in abstract Hilbert spaces is given.
In section 3 coherent state tomography is developed. In section 4, a
resolution of the unity is obtained which provides a relation between
Sudarshan's diagonal coherent state representation and Husimi-Kano $K-$%
function in the suggested tomographic setting. That identity resolution is
then recovered in the context of the generalized phase space distributions
associated with the Agarwal-Wolf $\Omega -$operator ordering \cite
{MandelWolf,AgWolf}. In section 5, generic deformations of coherent states
connected with $f-$oscillators, $q-$oscillators, and $s-$deformations
associated to operator ordering are developed. In section 6, photon number
tomography is reviewed and the new scheme of deformed tomography with both
nonlinearity and $s-$ordering is extended to photon number states. Some
conclusions and perspectives are drawn in section 7. An Appendix presents a
check of both the resolutions of the unity associated with the coherent
state and the photon number tomographies and a derivation of the Mehler's
formula in the holomorphic representation.

\section{Preliminaries}

In this section we give a short account of our previous works \cite
{PLA2,PalPalme}, in which we have given an interpretation of quantum
tomography in an abstract Hilbert space $\mathcal{H}$ by means of complete
sets of rank-one projectors $\left\{ P_{\mu }\right\} _{\mu \in M}$ , where $%
M$ is a set of (multi-) parameters, discrete or continuous, collectively
denoted by $\mu .$ In general, a tomogram of a quantum state $\left\vert
\psi \right\rangle $ is a positive real number $\mathcal{T}_{\psi }(\mu ),$
depending on the parameter $\mu $ which labels a set of states $\left\vert
\mu \right\rangle \in \mathcal{H},$ defined as 
\begin{equation}
\mathcal{T}_{\psi }(\mu ):=\left\vert \left\langle \mu |\psi \right\rangle
\right\vert ^{2}.
\end{equation}
Our main idea was to regard the tomogram $\mathcal{T}_{\psi }(\mu )$ as a
scalar product on the (Hilbert) space $\mathbb{H}$ of the rank-one
projectors $\left\vert \mu \right\rangle \left\langle \mu \right\vert
=P_{\mu }\rightarrow \left\vert P_{\mu }\right\rangle \in \mathbb{H}:$%
\begin{equation}
\mathcal{T}_{\psi }(\mu )=\mathrm{Tr}\left( P_{\mu }\rho _{\psi }\right)
=:\left\langle P_{\mu }|\rho _{\psi }\right\rangle .  \label{Tomo}
\end{equation}
Equation (\ref{Tomo}) may readily be used to define the tomogram of any
density operator $\hat{\rho}$ or any other (bounded) operator $\hat{A}$%
\begin{equation}
\mathcal{T}_{A}(\mu ):=\mathrm{Tr}\left( P_{\mu }\hat{A}\right)
=\left\langle P_{\mu }|A\right\rangle .  \label{Tomogen}
\end{equation}
Equation (\ref{Tomogen}) shows in general that to any operator $\hat{A}$ a
function $\left\langle \mu |\hat{A}|\mu \right\rangle $ of the variables $%
\mu $ corresponds in a given functional space. So, a tomograph may be
thought of as a de-quantization, and in fact is an useful tool to study the
quantum-classical transition by comparing classical limits of quantum
tomograms with the corresponding classical tomograms \cite{PLA1}. In the
same sense, the inverse correspondence $\mathcal{T}_{A}(\mu )\rightarrow 
\hat{A}$ may be considered to give a quantization. The reconstruction of the
operator $\hat{A}$ from its tomogram $\mathcal{T}_{A}(\mu )$ may be written
as 
\begin{equation}
\hat{A}=\sum\limits_{\mu \in M}\hat{G}_{\mu }\mathrm{Tr}\left( P_{\mu }\hat{A%
}\right) \Leftrightarrow \left\vert A\right\rangle =\sum\limits_{\mu \in
M}\left\vert G_{\mu }\right\rangle \left\langle P_{\mu }|A\right\rangle .
\end{equation}
In other words, the reconstruction of any operator is possible because the
tomographic set $\left\{ P_{\mu }\right\} _{\mu \in M}$ provides a
resolution of the identity (super-) operator on $\mathbb{H}$: 
\begin{equation}
\mathbb{\hat{I}}=\sum\limits_{\mu \in M}\hat{G}_{\mu }\mathrm{Tr}\left(
P_{\mu }\cdot \right) =\sum\limits_{\mu \in M}\left\vert G_{\mu
}\right\rangle \left\langle P_{\mu }\right\vert .
\end{equation}
We may then view $\left\vert G _{\mu }\right\rangle $ and$\ \left\langle
P_{\mu }\right\vert $ as dual supervectors. For instance for the spin
tomography, in the \textit{maximal} qu-bit case $M$ is the Bloch sphere $%
S^{2}$ of all rank-one projectors and for any operator $\hat{A}$\ holds \cite
{PLA2}: 
\begin{eqnarray}
\hat{A} &=&\int_{0}^{2\pi }\int_{0}^{\pi }\hat{G}(\theta ,\phi )\mathrm{Tr}%
(P(\theta ,\phi )A)\sin \theta d\theta d\phi  \label{A1} \\
&=&\int_{0}^{2\pi }\int_{0}^{\pi }P(\theta ,\phi )\mathrm{Tr}(\hat{G}(\theta
,\phi )A)\sin \theta d\theta d\phi ,  \label{A2}
\end{eqnarray}
where, in matrix form, 
\begin{equation*}
P(\theta ,\phi )=\frac{1}{2}\left[ 
\begin{array}{cc}
1+\cos \theta & e^{-i\phi }\sin \theta \\ 
e^{i\phi }\sin \theta & 1-\cos \theta
\end{array}
\right] ;~\hat{G}(\theta ,\phi )=\frac{1}{4\pi }\left[ 
\begin{array}{cc}
1+3\cos \theta & 3e^{-i\phi }\sin \theta \\ 
3e^{i\phi }\sin \theta & 1-3\cos \theta
\end{array}
\right] .
\end{equation*}

By definition, the set $\left\{ P_{\mu }\right\} _{\mu \in M}$ is \textit{%
tomographic} if it is complete in $\mathbb{H}$. A tomographic set determines
a \textit{tomograph} which is a functional, linear in the second argument 
\begin{equation}
(P_{\mu },\hat{A})\longrightarrow \mathcal{T}_{A}(\mu )=\mathrm{Tr}\left(
P_{\mu }\hat{A}\right) .  \label{Tomography}
\end{equation}
This definition is appropriate in the finite $n-$dimensional case, where 
\begin{equation*}
\left| \mu \right\rangle \in \mathcal{H}_{n}\Leftrightarrow P_{\mu }\in 
\mathbb{H}_{n^{2}}=B(\mathcal{H}_{n})=\mathcal{H}_{n}\otimes \mathcal{H}_{n},
\end{equation*}
but in the infinite dimensional case the relation $\mathbb{H}=B(\mathcal{H})$
is no more valid and there are several relevant spaces, as the space of
bounded operators $B(\mathcal{H})$ and that of compact operators $C(\mathcal{%
H})$, the space of Hilbert-Schmidt operators $\mathfrak{I}_{2}$ and that of
trace-class operators $\mathfrak{I}_{1}$. Their mutual relations are: 
\begin{equation*}
\mathfrak{I}_{1}\subset \mathfrak{I}_{2}\subset C(\mathcal{H})\subset B(%
\mathcal{H}).
\end{equation*}
$B(\mathcal{H})$ $($and $C\mathbb{(\mathcal{H}))}$ are Banach spaces, with
the norm $\left\| A\right\| =\sup_{(\left\| \psi \right\| =1)}\left\| A\psi
\right\| ,$ while $\mathfrak{I}_{2}$ is a Hilbert space with scalar product $%
\left\langle A|B\right\rangle =\mathrm{Tr}\left( A^{\dagger }B\right) $.
Finally, $\mathfrak{I}_{1}$ is a Banach space with the norm $\left\|
A\right\| _{1}=\mathrm{Tr}\left( \left| A\right| \right) .$ The following
inequalities hold true: 
\begin{equation*}
\left\| A\right\| \leq \left\| A\right\| _{2}\leq \left\| A\right\| _{1}.
\end{equation*}
So $\mathfrak{I}_{2}$, the only Hilbert space at our disposal to implement
our definition of tomographic set, is endowed with a topology which, when
restricted to the trace-class operators, is not equivalent to the topology
of $\mathfrak{I}_{1}$. This may have serious consequences. In fact, in the
finite dimensional case, the set $\left\{ P_{\mu }\right\} _{\mu \in M}$ is
complete iff 
\begin{equation}
\mathrm{Tr}\left( P_{\mu }A\right) =0\quad \forall \mu \in M\Longrightarrow
A=0.  \label{Azero}
\end{equation}
Such a condition guarantees the full reconstruction of any observable from
its tomograms. Now, in $\mathfrak{I}_{2}$, Eq. (\ref{Azero}) reads: 
\begin{equation}
\left\langle P_{\mu }|A\right\rangle =0\quad \forall \mu \in
M\Longrightarrow A=0\quad \&\quad A\in \mathfrak{I}_{2}.
\end{equation}
Then, as $\mathfrak{I}_{2}$ is a $\ast -$ideal in $B(\mathcal{H}),$ there
may exists a non-zero operator $B,$ which is bounded but not
Hilbert-Schmidt, such that 
\begin{equation*}
\mathrm{Tr}\left( P_{\mu }B\right) =0\quad \forall \mu \in M
\end{equation*}
In other words, different observables may be tomographically separated only
when their difference is Hilbert-Schmidt. Nevertheless there is a second
case, when the set $\left\{ P_{\mu }\right\} _{\mu \in M}$ of trace-class
operators is complete even in $\mathfrak{I}_{1}$. Then, recalling that $%
\mathfrak{I}_{1}$ is a $\ast -$ideal in its dual space $B(\mathcal{H})$: 
\begin{equation*}
\mathfrak{I}_{1}^{\ast }=B(\mathcal{H}),
\end{equation*}
the expression $\mathrm{Tr}\left( P_{\mu }A\right) $ is nothing but the
value of the linear functional $\mathrm{Tr}\left( \cdot A\right) $ in $%
P_{\mu }.$ Hence, Eq.(\ref{Azero}) holds unconditionally 
\begin{equation}
\mathrm{Tr}\left( P_{\mu }A\right) =0\quad \forall \mu \in M\Longrightarrow
0=\left\| \mathrm{Tr}\left( \cdot A\right) \right\| =\left\| A\right\|
\Longrightarrow A=0.
\end{equation}
Thus, the finest tomographies are those based on sets of rank-one projectors
which are complete both in $\mathfrak{I}_{2}$ and in $\mathfrak{I}_{1}$. As
a matter of fact, this is the case for the main tomographic sets, like the
symplectic, the photon number and the coherent state tomographic sets.

From a geometrical point of view, tomographic sets are \textquotedblleft
skew \textquotedblright\ sets of projectors. In other words, we define a set
of projectors to be \textit{skew} when it spans the whole Hilbert space.
Thus, any tomographic set is \textit{skew} as it is complete. Besides, we
define a set $\gamma $ of projectors to be \textit{locally skew }in $P_{0}$
, if it contains $P_{0}$ and any neighborhood of $P_{0}$ contains a skew
subset of $\gamma .$ Perhaps the simplest case of a tomographic set which is
skew and locally skew is provided by the coherent state tomographic set
discussed in the next sections.

\section{The coherent state tomography}

This tomographic set is generated by the displacement operators $\left\{ 
\mathcal{D}\left( z \right) \right\} $ depending on a complex parameter $z $%
\begin{equation}
\mathcal{D}\left( z \right) =\exp \left( z \hat{a}^{\dagger }-z ^{\ast }\hat{%
a}\right) ,\quad z \in \mathbb{C},
\end{equation}
which acting on the projector $\left\vert 0\right\rangle \left\langle
0\right\vert $ of the vacuum Fock state, $\hat{a}\left\vert 0\right\rangle
=0,$ yield the projectors 
\begin{equation}
\left\vert z \right\rangle \left\langle z \right\vert =\mathcal{D}\left( z
\right) \left\vert 0\right\rangle \left\langle 0\right\vert \mathcal{D}%
\left( z \right) ^{\dagger },\quad z \in \mathbb{C},
\end{equation}
associated to the usual coherent states 
\begin{equation*}
\left\vert z \right\rangle =\exp (-\frac{\left\vert z \right\vert ^{2}}{2}%
)\exp \left( z \hat{a}^{\dagger }\right) \exp \left( -z ^{\ast }\hat{a}%
\right) \left\vert 0\right\rangle =\exp (-\frac{\left\vert z \right\vert ^{2}%
}{2})\sum_{j=0}^{\infty }\frac{z ^{j}}{j!}\hat{a}^{\dagger j}\left\vert
0\right\rangle \,\ .
\end{equation*}

We recall that the coherent states are a (over-) complete set in the Hilbert
space $\mathcal{H}.$ Any bounded set containing a limit point $z_{0}$ in the
complex $z-$plane defines a complete set of coherent states containing a
limit point, the coherent state $\left\vert z_{0}\right\rangle ,$ in the
Hilbert space $\mathcal{H}$. In particular, any Cauchy sequence $\{z_{k}\}$
of complex numbers defines a Cauchy sequence of coherent states $\left\{
\left\vert z_{k}\right\rangle \right\} ,$ which is a complete set. The same
holds for any extracted subsequence. This completeness property holds as $%
\exp \left( \left\vert z\right\vert ^{2}/2\right) \left\langle z|\psi
\right\rangle $ is an entire analytic function of the complex variable $%
z^{\ast },$ for any $\left\vert \psi \right\rangle \in \mathcal{H}$. Then 
\begin{equation}
\left\langle z_{k}|\psi \right\rangle =0\quad \forall k\Rightarrow
\left\vert \psi \right\rangle =0,
\end{equation}
because $z_{0}^{\ast }$ is a non-isolated zero of an analytic function.

Besides, any bounded operator $A$ may be completely reconstructed from its
diagonal matrix elements $\left\langle z _{k}\left\vert A\right\vert z
_{k}\right\rangle .$ In fact, $\exp \left( \left\vert z \right\vert
^{2}/2+\left\vert z^{\prime}\right\vert ^{2}/2\right) \left\langle z
\left\vert A\right\vert z^{\prime}\right\rangle $ is an analytical function
of the complex variables $z ^{\ast },z^{\prime},$\ so it is uniquely
determined by its value $\exp \left( \left\vert z \right\vert ^{2}\right)
\left\langle z \left\vert A\right\vert z \right\rangle $ on the diagonal $%
z^{\prime}=z .$ This is an entire function of the real variables $\Re z ,\Im
z ,$ which is in turn uniquely determined by its values on any set with an
accumulation point.

The rank-one projectors associated to a complete set of coherent states are
complete in the Hilbert space $\mathfrak{I}_{2}$. In particular, any Cauchy
sequence $\left\{ \left\vert z _{k}\right\rangle \right\} $ generates a
tomographic set $\left\{ \left\vert z _{k}\right\rangle \left\langle z
_{k}\right\vert \right\} $. In fact, bearing in mind the previous remark on
the reconstruction of a bounded operator, it results 
\begin{equation}
\mathrm{Tr}(A\left\vert z _{k}\right\rangle \left\langle z _{k}\right\vert
)=\left\langle z _{k}\left\vert A\right\vert z _{k}\right\rangle =0\quad
\forall k\Rightarrow A=0\quad \&\quad A\in B(\mathcal{H}).
\end{equation}
This shows that a tomographic set of coherent state projectors is complete
even in $\mathfrak{I}_{1}.$ So it is skew. Moreover, any extracted
subsequence $\left\{ \left\vert z _{k_{n}}\right\rangle \left\langle z
_{k_{n}}\right\vert \right\} $ is again complete, so $\left\{ \left\vert z
_{k}\right\rangle \left\langle z _{k}\right\vert \right\} $ is locally skew
in its limit point. In the next section we address the case when $z$ varies
in the whole complex plane.

\section{The resolution of unity generated by the coherent state tomographic
set}

It is possible to interpret the well known Husimi-Kano $K$-symbol of a
(bounded) operator $\hat{A}$ as the coherent state (CS) tomogram of $\hat{A}$%
: 
\begin{equation}
K_{A}(z):=\left\langle z\left| \hat{A}\right| z\right\rangle =:\mathrm{Tr}%
(\left| z\right\rangle \left\langle z\right| \hat{A}).  \label{HK1}
\end{equation}
In particular, when $\hat{A}$ is chosen as a density operator $\hat{\rho},$
the identity holds 
\begin{equation}
\int \frac{d^{2}z}{\pi }\left\langle z\left| \hat{\rho}\right|
z\right\rangle =\mathrm{Tr}(\hat{\rho})=1,
\end{equation}
which allows for the probabilistic interpretation of the CS tomography. As a
matter of fact \cite{SudKlaubook} the $K$-symbol exists also for a number of
non-bounded operators. The CS tomographic set is complete both in $%
\mathfrak{I}_{2}$, the space of Hilbert-Schmidt operators, and in $%
\mathfrak{I}_{1},$ the space of trace class operators acting on the space of
states. In fact, the formulae 
\begin{equation}
\hat{A}=\int \frac{d^{2}z}{\pi }\frac{d^{2}z^{\prime }}{\pi }\left\langle
z\left| \hat{A}\right| z^{\prime }\right\rangle \left| z\right\rangle
\left\langle z^{\prime }\right|
\end{equation}
and \cite{MandelWolf} 
\begin{equation}
\left\langle z\left| \hat{A}\right| z^{\prime }\right\rangle =\mathrm{e}^{-%
\frac{\left| z\right| ^{2}+\left| z^{\prime }\right| ^{2}}{2}%
}\sum_{n,m=0}^{\infty }\frac{(z^{\ast })^{n}(z^{\prime })^{m}}{n!m!}\left[ 
\frac{\partial ^{n+m}}{\partial z^{\ast n}\partial z^{m}}\left( \mathrm{e}%
^{\left| z\right| ^{2}}\left\langle z\left| \hat{A}\right| z\right\rangle
\right) \right] _{_{z=0}^{z^{\ast }=0}}  \label{19}
\end{equation}
show that if the tomograms $\left\langle z\left| \hat{A}\right|
z\right\rangle $ of a bounded operator $\hat{A}$ vanish for any $z\in 
\mathbb{C}$ , then $\hat{A}$ is the zero operator. Equation (\ref{19}) is
implicit in Eq. (6) of Sudarshan Ref. \cite{Sud63}.

So, a resolution of the unity exists, which allows for the full
reconstruction of any (bounded) operator from its CS tomograms. We are
interested in the explicit determination of such a formula. Now, the
Sudarshan's diagonal coherent state representation $\phi_{A}(z)$ of an
operator $\hat{A}$ is defined through the equation 
\begin{equation}
\hat{A}=\int \frac{d^{2}z}{\pi }\phi_{A}(z)\left| z\right\rangle
\left\langle z\right| .  \label{DefQ}
\end{equation}
Bearing in mind the previous qu-bit reconstruction formula, Eq.(\ref{A2}),
we may note the analogy with Eq.(\ref{DefQ}) defining $\phi_{A}(z)$. Then,
if our guess is right, we need to get explicitly the first form of the
reconstruction formula, Eq.(\ref{A1}). In other words, we have to invert the
well-known relation: 
\begin{equation}
K_{A}(z^{\prime })=\left\langle z^{\prime }\left| \hat{A}\right| z^{\prime
}\right\rangle =\int \frac{d^{2}z}{\pi }\phi_{A}(z)\left| \left\langle
z|z^{\prime }\right\rangle \right| ^{2}=\int \frac{d^{2}z}{\pi }\phi_{A}(z)%
\mathrm{e}^{-\left| z-z^{\prime }\right| ^{2}}  \label{Fourtrans}
\end{equation}
which follows at once from Eq.(\ref{DefQ}) defining $\phi_{A}(z)$. This
relation shows that $K_{A}(z^{\prime })$ is given by the convolution product
of $\phi_{A}$ times a gaussian function. Then, denoting with $%
K_{A}(z_{R}^{\prime },z_{I}^{\prime })$ and $\phi_{A}(z_{R},z_{I})$ the $K$
and $\phi$ symbols, with $z^{\prime }=z_{R}^{\prime }+iz_{I}^{\prime }$ ; $%
z=z_{R}+iz_{I}$, the Fourier transform \cite{Schleich} of Eq. (\ref
{Fourtrans}) reads: 
\begin{eqnarray}
&&\int \frac{dz_{R}^{\prime }dz_{I}^{\prime }}{2\pi }K_{A}(z_{R}^{\prime
},z_{I}^{\prime })\mathrm{e}^{-i(\xi z_{R}^{\prime }+\eta z_{I}^{\prime })}=
\\
&&\int \frac{dz_{R}^{\prime }dz_{I}^{\prime }}{2\pi }\int \frac{dz_{R}dz_{I}%
}{\pi }\phi_{A}(z_{R},z_{I})\mathrm{e}^{-\left( z_{R}-z_{R}^{\prime }\right)
^{2}-\left( z_{I}-z_{I}^{\prime }\right) ^{2}}\mathrm{e}^{-i(\xi
z_{R}^{\prime }+\eta z_{I}^{\prime })}  \notag
\end{eqnarray}
and we readily obtain 
\begin{equation}
\tilde{K}_{A}(\xi ,\eta )=\mathrm{e}^{-(\xi ^{2}+\eta ^{2})/4}\tilde{\phi}%
_{A}(\xi ,\eta ),
\end{equation}
from which 
\begin{equation}
\tilde{\phi}_{A}(\xi ,\eta )=\mathrm{e}^{(\xi ^{2}+\eta ^{2})/4}\tilde{K}%
_{A}(\xi ,\eta ),
\end{equation}
that formally yields 
\begin{equation}
\phi_{A}(z_{R},z_{I})=\int \frac{d\xi d\eta }{2\pi }\mathrm{e}^{(\xi
^{2}+\eta ^{2})/4}\tilde{K}_{A}(\xi ,\eta )\mathrm{e}^{i(\xi z_{R}+\eta
z_{I})}.
\end{equation}
The presence of the anti-gaussian factor shows that the inverse Fourier
transform of $\tilde{\phi}_{A}(\xi ,\eta )$ exists only when the asymptotic
decay of $\tilde{K}_{A}(\xi ,\eta )$ is faster than the growth of $\mathrm{e}%
^{(\xi ^{2}+\eta ^{2})/4}.$ However, the integral always exists as a
distribution, as was proven in Ref.\cite{MehtaSud65}. By virtue of this
remark, we may go on and substitute the previous expression into Eq.(\ref
{DefQ}) getting 
\begin{eqnarray}
&&\hat{A}=\int \frac{d^{2}z}{\pi }\left[ \int \frac{d\xi d\eta }{2\pi }%
\mathrm{e}^{(\xi ^{2}+\eta ^{2})/4}\tilde{K}_{A}(\xi ,\eta )\mathrm{e}%
^{i(\xi z_{R}+\eta z_{I})}\right] \left| z\right\rangle \left\langle
z\right| = \\
\int \frac{d^{2}z}{\pi } &&\left[ \int \frac{d\xi d\eta }{2\pi }\int \frac{%
dz_{R}^{\prime }dz_{I}^{\prime }}{2\pi }K_{A}(z_{R}^{\prime },z_{I}^{\prime
})\mathrm{e}^{(\xi ^{2}+\eta ^{2})/4}\mathrm{e}^{i\left[ \xi \left(
z_{R}-z_{R}^{\prime }\right) +\eta \left( z_{I}-z_{I}^{\prime }\right) %
\right] }\right] \left| z\right\rangle \left\langle z\right| .  \notag
\end{eqnarray}
Upon interchanging the order of integration, we may write the expected
reconstruction formula as 
\begin{equation}
\hat{A}=\int \frac{d^{2}z^{\prime }}{\pi }\hat{G}(z^{\prime
})K_{A}(z^{\prime }),  \label{Areconstr}
\end{equation}
where the Gram-Schmidt operator $\hat{G}(z^{\prime })$ reads: 
\begin{equation}
\hat{G}(z^{\prime }):=\int \frac{d^{2}z}{2\pi }\int \frac{d\xi d\eta }{2\pi }%
\mathrm{e}^{(\xi ^{2}+\eta ^{2})/4}\mathrm{e}^{i\left[ \xi \left(
z_{R}-z_{R}^{\prime }\right) +\eta \left( z_{I}-z_{I}^{\prime }\right) %
\right] }\left| z\right\rangle \left\langle z\right| .  \label{GramCS}
\end{equation}
In other words, the resolution of the unity generated by the CS tomographic
set is 
\begin{equation}
\mathbb{\hat{I}}=\int \frac{d^{2}z^{\prime }}{\pi }\hat{G}(z^{\prime })%
\mathrm{Tr}(\left| z^{\prime }\right\rangle \left\langle z^{\prime }\right|
\cdot )\quad .
\end{equation}

We observe that 
\begin{eqnarray}
&&K_{{G}(z^{\prime })}(z)=\left\langle z\left\vert \hat{G}(z^{\prime
})\right\vert z\right\rangle =\int \frac{d\xi d\eta }{2\pi }\mathrm{e}^{i%
\left[ -\xi z_{R}^{\prime }-\eta z_{I}^{\prime }\right] }\mathrm{e}^{(\xi
^{2}+\eta ^{2})/4}\tilde{K}_{\left\vert z\right\rangle \left\langle
z\right\vert }\left( \xi ,\eta \right)  \notag \\
&=&\int \frac{d\xi d\eta }{2\pi }\mathrm{e}^{i\left[ -\xi z_{R}^{\prime
}-\eta z_{I}^{\prime }\right] }\tilde{\phi}_{\left\vert z\right\rangle
\left\langle z\right\vert }\left( \xi ,\eta \right) =\phi_{\left\vert
z\right\rangle \left\langle z\right\vert }\left( z^{\prime }\right) .
\label{K_Q}
\end{eqnarray}
This, substituted in the reconstruction formula Eq.(\ref{Areconstr}), gives
also the reproducing kernel formula 
\begin{equation}
K_{{G}(z^{\prime \prime })}(z)=\int \frac{d^{2}z^{\prime }}{\pi }%
\left\langle z\left\vert \hat{G}(z^{\prime })\right\vert z\right\rangle K_{{G%
}(z^{\prime \prime })}(z^{\prime })=\int \frac{d^{2}z^{\prime }}{\pi }K_{{G}%
(z^{\prime })}(z)K_{{G}(z^{\prime \prime })}(z^{\prime })
\end{equation}
or equivalently 
\begin{equation}
\phi_{\left\vert z\right\rangle \left\langle z\right\vert }\left( z^{\prime
\prime }\right) =\int \frac{d^{2}z^{\prime }}{\pi }\phi_{\left\vert
z\right\rangle \left\langle z\right\vert }\left( z^{\prime }\right)
\phi_{\left\vert z^{\prime }\right\rangle \left\langle z^{\prime
}\right\vert }\left( z^{\prime \prime }\right) \ .
\end{equation}
Since 
\begin{equation}
\left\vert z\right\rangle \left\langle z\right\vert =\int \frac{%
d^{2}z^{\prime }}{\pi }\phi_{\left\vert z\right\rangle \left\langle
z\right\vert }\left( z^{\prime }\right) \left\vert z^{\prime }\right\rangle
\left\langle z^{\prime }\right\vert \Leftrightarrow \phi_{\left\vert
z\right\rangle \left\langle z\right\vert }\left( z^{\prime }\right) =\pi
\delta (z-z^{\prime })~,
\end{equation}
we remark that Eq.(\ref{K_Q}) amounts to the orthonormality relations 
\begin{equation}
K_{{G}(z^{\prime })}(z)=\mathrm{Tr}(\left\vert z\right\rangle \left\langle
z\right\vert {\hat{G}}(z^{\prime }))=\pi \delta (z-z^{\prime })
\end{equation}
between dual sets of supervectors.

In view of Eq.(\ref{19}), we can use the previous expression of $K_{G}$ to
check the reconstruction formula, Eq.(\ref{Areconstr}), in matrix form as: 
\begin{equation}
\left\langle z\left| \hat{A}\right| z\right\rangle =\int \frac{%
d^{2}z^{\prime }}{\pi }K_{{G}(z^{\prime })}(z)K_{A}(z^{\prime })=K_{A}(z).
\end{equation}
A check of the reconstruction formula in the position representation is
given in the Appendix.

In view of the generalization of our results in the next section, it is
expedient to make preliminarily contact with the generalized phase space
distributions associated with the Agarwal-Wolf $\Omega -$operator ordering 
\cite{MandelWolf,AgWolf}. In fact, in the context of that operator ordering
theory, the $K$ and $\phi$ symbols, appearing in our direct and dual
reconstruction formulae, are related to the Wick (i.e., normal) and
anti-Wick ordering respectively. As it is well known, it is possible to
express any operator $A(\hat{a},\hat{a}^{\dagger })$ in terms of a function $%
F_{A}(z,z^{\ast })$ of two complex variables $u=z,v=z^{\ast },$ by means of
an operator $\hat{\Omega}$ which may be realized as integral operator 
\begin{equation}
A(\hat{a},\hat{a}^{\dagger })=\int \frac{d^{2}z}{\pi }F_{A}(z,z^{\ast
})\Delta ^{(\Omega )}(z-\hat{a},z^{\ast }-\hat{a}^{\dagger })  \label{AW1}
\end{equation}
where the $\Omega -$ ordered delta function 
\begin{equation}
\Delta ^{(\Omega )}(z-\hat{a},z^{\ast }-\hat{a}^{\dagger })=\int \frac{d^{2}w%
}{\pi }\Omega (w,w^{\ast })\exp (w(z^{\ast }-\hat{a}^{\dagger })-w^{\ast }(z-%
\hat{a}))  \label{AW2}
\end{equation}
is obtained by the displacement operator and an entire analytic function $%
\Omega (w,w^{\ast })$ of two complex variables, which characterizes the
particular rule of association between $\hat{A}$ and $F_{A}.$ For instance,
the Wick ordered delta function $\Delta ^{(W)}$ is given by $\Omega
(w,w^{\ast })=\exp (ww^{\ast }/2),$ while the anti-Wick ordered $\Delta ^{(%
\bar{W})}$ is given by the reciprocal function $\exp (-ww^{\ast }/2).$

The function $F_{A}$ is associated to the operator $\hat{A}$ \textit{via}
the anti-$\Omega -$ ordered delta function: 
\begin{equation}
F_{A}(z,z^{\ast })=\mathrm{Tr}[A(\hat{a},\hat{a}^{\dagger })\Delta ^{(\bar{%
\Omega})}(z-\hat{a},z^{\ast }-\hat{a}^{\dagger })].  \label{AW3}
\end{equation}

It is apparent the similarity between Eq. (\ref{AW1}) and our previous
reconstruction formulae, but Agarwal-Wolf formulae, in general, do not
contain rank-one projectors, which instead characterize our tomographic
theory. However, it results that the anti-Wick ordered delta function is a
rank-one projector: 
\begin{equation}
\Delta ^{(\bar{W})}(z-\hat{a},z^{\ast }-\hat{a}^{\dagger })=\left|
z\right\rangle \left\langle z\right| ,
\end{equation}
so that the Wick ordered delta function $\Delta ^{(W)}$ is nothing but our
Gram-Schmidt operator $\hat{G}$ of Eq.(\ref{GramCS}), and Eq.(\ref{AW3})
yields the Husimi -Kano function $K_{A}$ while Eq.(\ref{AW1}) is just our
direct reconstruction formula Eq.(\ref{Areconstr}). When exchanging the role
of the two orderings one obtains respectively the Sudarshan's diagonal
representation $\phi _{A}$ and the dual form of our reconstruction formula,
Eq.(\ref{DefQ}).

\section{The $\left( f,s\right) -$deformed CS tomographic set}

The usual coherent states, that we have recast in a tomographic setting in
the previous sections, date back to Schr\"{o}dinger. We have defined them by
applying the unitary family of the displacement operators $\{\mathcal{D}%
\left( z\right) \}$ on the vacuum state $\left\vert 0\right\rangle ,$ such
that $\hat{a}\left\vert 0\right\rangle =0.$ Recently, a number of so-called
generalized coherent states (GCS) have been introduced, in connection with
oscillator algebras different from the standard one. In particular, an
interesting family of such GCS, related with deformed oscillator algebras,
is the set of \textit{nonlinear }CS \cite{MVogel,Sudf,ZAC2000} which arises
from a deformation ruled by a class $\{f\left( \hat{n}\right) \}$ of
functions of the number operator. This class is the set of real functions
such that 
\begin{equation}
0<c<\left[ f\left( n\right) \right] !<c^{\prime }<\infty ,\,\ n\in \mathbb{N}%
;\,\ \left[ f\left( n\right) \right] !:=f\left( n\right) f\left( n-1\right)
...f\left( 1\right) .
\end{equation}
Following \cite{ZAC2000}, that we briefly recall mainly to establish
notation, we assume that these functions satisfy some further condition:
they are continuous and moreover 
\begin{equation}
f\left( n\right) =1,\mathbb{\quad }n=0,-1,-2,... \quad .
\end{equation}
Now, introducing the selfadjoint $f-$deformation operator 
\begin{equation}
\mathcal{E}_{f}:=f\left( \hat{n}\right) \prod\limits_{m=1}^{\infty }f\left( 
\hat{n}-m\right) =:\left[ f\left( \hat{n}\right) \right] !
\end{equation}
we have 
\begin{equation}
\mathcal{E}_{f}^{-1}\hat{a}\mathcal{E}_{f}=\hat{a}f\left( \hat{n}\right) =:%
\hat{A};\mathbb{\quad }\mathcal{E}_{f}^{-1}\hat{a}^{\dagger }\mathcal{E}_{f}=%
\frac{1}{f\left( \hat{n}\right) }\hat{a}^{\dagger }=:\hat{A}_{f}^{\dagger }
\quad .
\end{equation}
Then, the $f-$deformed displacement operator is 
\begin{equation}
\mathcal{D}_{f}\left( z\right) =\mathcal{E}_{f}^{-1}\mathcal{D}\left(
z\right) \mathcal{E}_{f}=\exp \left( z\hat{A}_{f}^{\dagger }-z^{\ast }\hat{A}%
\right) .
\end{equation}
Note that $\hat{A}_{f}^{\dagger }$ appears instead of $\hat{A}^{\dagger
}=f\left( \hat{n}\right) \hat{a}^{\dag }$ in the deformed displacement
operator. This $\hat{A}_{f}^{\dagger }$ obeys the commutation rule $\left[ 
\hat{A},\hat{A}_{f}^{\dagger }\right] =\left[ \hat{a},\hat{a}^{\dagger }%
\right] =1.$

Now, it is possible to compose the $f-$deformation with the $s-$deformation
of operator ordering \cite{SudBrandeis,Cahill-Glauber}, so defining the $%
\left( f,s\right) -$deformed displacement operators as: 
\begin{equation}
\mathcal{D}_{f,s}\left( z\right) :=\exp \left( \frac{1}{2}szz^{\ast }\right) 
\mathcal{D}_{f}\left( z\right) =\exp \left( \frac{1}{2}szz^{\ast }\right)
\exp \left( z\hat{A}_{f}^{\dagger }-z^{\ast }\hat{A}\right) ,
\end{equation}
where $z\in \mathbb{C},\,\ -1\leq s\leq 1.$

The $\left( f,s\right) -$deformed displacement operators, acting on the
vacuum, generate the $\left( f,s\right) -$deformed coherent states: 
\begin{eqnarray}
&&\left\vert z ;f,s\right\rangle :=\mathrm{e}^{\left( 1+s\right) \left\vert
z \right\vert ^{2}/2}N_{z ,f}\mathcal{D}_{f}\left( z \right) \left\vert
0\right\rangle =\mathrm{e}^{\left( 1+s\right) \left\vert z \right\vert
^{2}/2}N_{z ,f}\mathcal{E}_{f}^{-1}\mathcal{D}\left( z \right) \mathcal{E}%
_{f}\left\vert 0\right\rangle  \notag \\
&&=\mathrm{e}^{\left( 1+s\right) \left\vert z \right\vert ^{2}/2}N_{z ,f}%
\mathcal{E}_{f}^{-1}\mathcal{D}\left( z \right) f\left( 0\right) \left\vert
0\right\rangle =\mathrm{e}^{\left( 1+s\right) \left\vert z \right\vert
^{2}/2}N_{z ,f}\mathcal{E}_{f}^{-1}\left\vert z \right\rangle
\end{eqnarray}
where the normalizing factor is 
\begin{equation}
N_{z ,f}=\left[ \sum\limits_{k=0}^{\infty }\frac{\left\vert z \right\vert
^{2k}}{k!\left( \left[ f\left( k\right) \right] !\right) ^{2}}\right]
^{-1/2}.  \label{NormNdef}
\end{equation}

Now, we are ready for the tomographic setting of such deformed states. We
define the $\left( f,s\right) -$deformed $K-$function of an operator $\hat{B}
$ as: 
\begin{eqnarray}
&&K_{B}^{f,s}\left( z \right) :=\left\langle z ;f,s\left\vert \hat{B}%
\right\vert z ;f,s\right\rangle = \\
&&\left\langle z \left\vert \mathrm{e}^{\left( 1+s\right) \left\vert z
\right\vert ^{2}/2}N_{z ,f}\mathcal{E}_{f}^{-1}\hat{B}\mathrm{e}^{\left(
1+s\right) \left\vert z \right\vert ^{2}/2}N_{z ,f}\mathcal{E}%
_{f}^{-1}\right\vert z \right\rangle =:\mathrm{e}^{\left( 1+s\right)
\left\vert z \right\vert ^{2}}N_{z ,f}^{2}K_{B\left( f\right) }\left( z
\right) ,  \notag
\end{eqnarray}
where 
\begin{equation}
\hat{B}\left( f\right) := \mathcal{E}_{f}^{-1}\hat{B}\mathcal{E}_{f}^{-1}.
\end{equation}
So, the $\left( f,s\right) -$deformed $K-$function of an operator $\hat{B}$
is the $K-$function of the $f-$deformed operator $\hat{B}\left( f\right) $
times a known $c-$number.

Thus, from the reconstruction formula of the non-deformed CS tomography: 
\begin{equation}
\hat{B}\left( f\right) =\int \frac{d^{2}z }{\pi }\hat{G}(z )K_{B\left(
f\right) }(z ),
\end{equation}
we get the reconstruction formula for the $\left( f,s\right) -$deformed CS
tomography: 
\begin{eqnarray}
\hat{B} &=&\int \frac{d^{2}z }{\pi }\mathcal{E}_{f}\hat{G}(z )\mathcal{E}%
_{f}K_{B\left( f\right) }(z )=\int \frac{d^{2}z }{\pi }\left[ \mathcal{E}_{f}%
\hat{G}(z )\mathcal{E}_{f}\mathrm{e}^{-\left( 1+s\right) \left\vert z
\right\vert ^{2}}N_{z ,f}^{-2}\right] K_{B}^{f,s}(z )  \notag \\
&=&:\int \frac{d^{2}z }{\pi }\hat{G}^{f,s}(z )K_{B}^{f,s}(z ).
\label{reconstrudef}
\end{eqnarray}
Remembering the expression of $\hat{G}$, Eq.(\ref{GramCS}), we have
explicitly the $\left( f,s\right) -$deformed Gram-Schmidt operator $\hat{G}%
^{f,s}$ as: 
\begin{eqnarray*}
&&\hat{G}^{f,s}(z ) =\frac{\mathrm{e}^{-\left( 1+s\right) \left\vert z
\right\vert ^{2}}}{N_{z ,f}^{2}}\int \frac{d^{2}z^{\prime}}{2\pi }\int \frac{%
d\xi d\eta }{2\pi }\mathrm{e}^{(\xi ^{2}+\eta ^{2})/4}\mathrm{e}^{i\left[
\xi \left( z^{\prime}_{R}-z _{R}\right) +\eta \left( z^{\prime}_{I}-z
_{I}\right) \right] }\mathcal{E}_{f}\left\vert z^{\prime}\right\rangle
\left\langle z^{\prime}\right\vert \mathcal{E}_{f} \\
=&&\frac{\mathrm{e}^{-2\left( 1+s\right) \left\vert z \right\vert ^{2}}}{%
N_{z ,f}^{4}}\int \frac{d^{2}z^{\prime}}{2\pi }\int \frac{d\xi d\eta }{2\pi }%
\mathrm{e}^{(\xi ^{2}+\eta ^{2})/4}\mathrm{e}^{i\left[ \xi \left(
z^{\prime}_{R}-z _{R}\right) +\eta \left( z^{\prime}_{I}-z _{I}\right) %
\right] }\mathcal{E}_{f}^{2}\left\vert z^{\prime};f,s\right\rangle
\left\langle z^{\prime};f,s\right\vert \mathcal{E}_{f}^{2}
\end{eqnarray*}

We recall \cite{ZAC2000} that $\mathcal{E}_{f}$ is selfadjoint and moreover $%
\mathcal{E}_{f}\mathcal{E}_{f}^{\dagger }=\mathcal{E}_{f}^{2}$ is a bounded
positive operator (with bounded inverse) which satisfies the intertwining
condition: 
\begin{equation}
\hat{A}^{\dag }\mathcal{E}_{f}^{2}=\mathcal{E}_{f}^{2}\hat{A}_{f}^{\dag }~.
\end{equation}
In other words, the operator $\hat{A}_{f}^{\dag }$ is nothing but the
adjoint of the $f-$deformed operator $\hat{A}$ with respect to a new scalar
product in our carrier Hilbert space $\mathcal{H}$: 
\begin{equation}
\left( \varphi ,\psi \right) _{f}:=\left( \varphi ,\mathcal{E}_{f}^{2}\psi
\right) \Rightarrow \left( \hat{A}\varphi ,\psi \right) _{f}=\left( \varphi ,%
\hat{A}^{\dag }\mathcal{E}_{f}^{2}\psi \right) =\left( \varphi ,\hat{A}%
_{f}^{\dag }\psi \right) _{f}
\end{equation}

So, a better notation would be $\hat{A}^{\dag _{f}}$ instead of $\hat{A}%
_{f}^{\dag }\ ,$ however we shall use the latter one.

The presence, in the expression of the deformed Gram-Schmidt operator, of
the deformed projector 
\begin{equation}
\mathcal{E}_{f}^{2}\left\vert z^{\prime };f,s\right\rangle \left\langle
z^{\prime };f,s\right\vert \mathcal{E}_{f}^{2}
\end{equation}
is reminiscent of the new deformed scalar product, and allows the
representation: 
\begin{eqnarray}
&&\left\langle w;f,s\left\vert \hat{G}^{f,s}(z)\right\vert w^{\prime
};f,s\right\rangle =\frac{\mathrm{e}^{-2\left( 1+s\right) \left\vert
z\right\vert ^{2}}}{N_{z,f}^{4}}\times \\
&&\int \frac{d^{2}z^{\prime }}{2\pi }\int \frac{d\xi d\eta }{2\pi }\mathrm{e}%
^{(\xi ^{2}+\eta ^{2})/4}\mathrm{e}^{i\left[ \xi \left( z_{R}^{\prime
}-z_{R}\right) +\eta \left( z_{I}^{\prime }-z_{I}\right) \right]
}\left\langle w;f,s|z^{\prime };f,s\right\rangle _{f}\left\langle z^{\prime
};f,s|w^{\prime };f,s\right\rangle _{f}\ .  \notag
\end{eqnarray}
When $w^{\prime }=w,$ we get 
\begin{equation}
\left[ K_{\left\vert w;f,s\right\rangle \left\langle w;f,s\right\vert }^{f,s}%
\right] _{f}\left( z^{\prime }\right) :=\left\langle w;f,s|z^{\prime
};f,s\right\rangle _{f}\left\langle z^{\prime };f,s|w;f,s\right\rangle
_{f}=K_{\mathcal{E}_{f}^{2}\left\vert w;f,s\right\rangle \left\langle
w;f,s\right\vert \mathcal{E}_{f}^{2}}^{f,s}\left( z^{\prime }\right)
\end{equation}
and 
\begin{eqnarray}
K_{G^{f,s}(z)}^{f,s}\left( w\right) &=&\left\langle w;f,s\left\vert \hat{G}%
^{f,s}(z)\right\vert w;f,s\right\rangle  \notag \\
=\frac{\mathrm{e}^{-2\left( 1+s\right) \left\vert z\right\vert ^{2}}}{%
N_{z,f}^{4}} &&\int \frac{d\xi d\eta }{2\pi }\mathrm{e}^{i\left[ -\xi
z_{R}-\eta z_{I}\right] }\mathrm{e}^{(\xi ^{2}+\eta ^{2})/4}\tilde{K}_{%
\mathcal{E}_{f}^{2}\left\vert w;f,s\right\rangle \left\langle
w;f,s\right\vert \mathcal{E}_{f}^{2}}^{f,s}\left( \xi ,\eta \right)  \notag
\\
=\frac{\mathrm{e}^{-2\left( 1+s\right) \left\vert z\right\vert ^{2}}}{%
N_{z,f}^{4}} &&\phi_{\mathcal{E}_{f}^{2}\left\vert w;f,s\right\rangle
\left\langle w;f,s\right\vert \mathcal{E}_{f}^{2}}^{f,s}\left( z\right) ,
\end{eqnarray}
which may be compared with the analogous formula of the standard CS case,
Eq.(\ref{K_Q}).

Let us consider now an important case of particular deformations, the $q-$
deformations. They are defined by considering the so called $q-$ oscillators 
\cite{Bied,Mc}. For $q-$ oscillators, the analog of annihilation and
creation operators $\hat{a},\hat{a}^{\dag }$ are the operators $\hat{a}_{q}$
and $\hat{a}_{q}^{\dag }$ which obey the relation: 
\begin{equation}
\hat{a}_{q}\hat{a}_{q}^{\dag }-q\hat{a}_{q}^{\dag }\hat{a}_{q}=q^{\hat{n}%
},\quad \hat{n}=\hat{a}^{\dag }\hat{a},  \label{q1}
\end{equation}
where $q$ is a $c-$number. When $q\rightarrow 1,$ the relation of Eq.(\ref
{q1}) becomes the standard boson commutation relation. One can introduce $q-$%
coherent states studying the eigenvalue problem 
\begin{equation}
\hat{a}_{q}\left| z ;q\right\rangle =z \left| z ;q\right\rangle\,\ .
\label{q2}
\end{equation}
The properties of $q-$oscillators and $q-$coherent states were intensively
discussed in the literature. As it was shown, e.g., in Ref.\cite{Sudf}, the $%
q-$deformations are a particular case of $f-$deformations, with a specific
function $f_q$ coding the $q-$nonlinearity of the $q-$oscillator vibrations,
which reads 
\begin{equation}
f_{q}( \hat{n} ) := \sqrt{\frac{\sinh \lambda \hat{n}}{\lambda \hat{n}}},
\quad q:=\text{e}^{\lambda}\,\ .
\end{equation}
One has: 
\begin{equation}
\hat{a}_{q}=\hat{a} f_{q}( \hat{n} ).  \label{q3}
\end{equation}
The operator $\hat{a}_{q}$ of Eq.(\ref{q3}) and its adjoint $\hat{a}%
_{q}^{\dag}$ satisfy the relation of Eq.(\ref{q1}). As a result, all the
previous formulae for the deformed coherent states are valid even for $q-$%
deformations. In particular, this holds for the reconstruction formula Eq.(%
\ref{reconstrudef}) and the associated resolution of the unity, in which we
have only to replace $f$ by $f_q$.

\section{Photon number tomographies}

\subsection{The photon number tomographic set}

The photon number tomography is generated by the irreducible family $\{\hat{a%
}^{\dagger }\hat{a},\left\{ \mathcal{D}\left( z\right) \right\} \}$, where
the displacement operators $\left\{ \mathcal{D}\left( z\right) \right\} $
act on the number operator $\hat{a}^{\dagger }\hat{a}$ of the harmonic
oscillator, yielding the family of selfadjoint operators $\hat{T}(z):$ 
\begin{equation}
\hat{T}(z)=\mathcal{D}\left( z\right) \hat{a}^{\dagger }\hat{a}\mathcal{D}%
\left( z\right) ^{\dagger },\quad z\in \mathbb{C}
\end{equation}
which has the spectrum of $\hat{a}^{\dagger }\hat{a},$ eigenvectors $%
\left\vert nz\right\rangle =\mathcal{D}\left( z\right) \left\vert
n\right\rangle $ and associated projectors $\left\vert nz\right\rangle
\left\langle nz\right\vert $.

We recall that the photon number projectors' set, containing the complete
set of the coherent state projectors, is a tomographic set complete both in $%
\mathfrak{I}_{2}$ and $\mathfrak{I}_{1}.$ For the same reason, any Cauchy
sequence $\left\{ \left\vert nz _{k}\right\rangle \left\langle nz
_{k}\right\vert \right\} $ is locally skew in its limit point \cite{PalPalme}%
.

The whole set of photon number projectors generates the resolution of the
unity: 
\begin{equation}
\mathbb{\hat{I}}=\sum\limits_{n=0}^{\infty }\int \frac{d^{2}z}{\pi }\hat{G}%
_{ \lambda }\left( n,z\right) \mathrm{Tr}(\left\vert nz\right\rangle
\left\langle nz\right\vert \cdot )\quad .  \label{Photoris}
\end{equation}
The Gram-Schmidt operator $\hat{G}_{ \lambda }$ is given by 
\begin{equation}
\hat{G}_{ \lambda }\left( n,z\right) =\frac{4}{1-\lambda ^{2}}\left( \frac{%
\lambda +1}{\lambda -1}\right) ^{n}\mathcal{D}\left( z\right) \left( \frac{%
\lambda -1}{\lambda +1}\right) ^{\hat{a}^{\dagger }\hat{a}}\mathcal{D}%
^{\dagger }\left( z\right) .  \label{Phonumk}
\end{equation}
Here $\lambda $ is a real parameter, $-1<\lambda <1,$ which labels the
family of equivalent kernels $\hat{G}_{ \lambda }\left( n,z\right) .$

The check of the matrix form of the resolution of the unity, Eq.(\ref
{Photoris}), in the position representation is given in the Appendix.

\subsection{The $\left( f,s\right) -$deformed photon number tomographic set}

We can rephrase the section on the $\left( f,s\right) -$deformed coherent
states, by letting the $\left( f,s\right) -$deformed displacement operators
act on the eigenstates $\left\vert n\right\rangle $ of the harmonic
oscillator, so generating the $\left( f,s\right) -$deformed photon number
states: 
\begin{eqnarray}
\left\vert nz ;f,s\right\rangle &:&=\mathrm{e}^{\left( 1+s\right) \left\vert
z \right\vert ^{2}/2}N_{z ,f}\mathcal{D}_{f}\left( z \right) \left\vert
n\right\rangle =\mathrm{e}^{\left( 1+s\right) \left\vert z \right\vert
^{2}/2}N_{z ,f}\mathcal{E}_{f}^{-1}\mathcal{D}\left( z \right) \mathcal{E}%
_{f}\left\vert n\right\rangle \\
&=&\mathrm{e}^{\left( 1+s\right) \left\vert z \right\vert ^{2}/2}N_{z ,f}%
\mathcal{E}_{f}^{-1}\mathcal{D}\left( z \right) \left[ f\left( n\right) %
\right] !\left\vert n\right\rangle =\mathrm{e}^{\left( 1+s\right) \left\vert
z \right\vert ^{2}/2}N_{z ,f}\left[ f\left( n\right) \right] !\mathcal{E}%
_{f}^{-1}\left\vert nz \right\rangle  \notag
\end{eqnarray}
where the normalizing factor $N_{z ,f}$\ is again given by Eq.(\ref{NormNdef}%
).

Now, we define the $\left( f,s\right) -$deformed $K-$function of an operator 
$\hat{B}$ as: 
\begin{eqnarray}
K_{B}^{f,s}\left( nz\right) &:=&\left\langle nz;f,s\left\vert \hat{B}%
\right\vert nz;f,s\right\rangle \\
&=&\mathrm{e}^{\left( 1+s\right) \left\vert z\right\vert ^{2}}N_{z,f}^{2}%
\left[ f^{2}\left( n\right) \right] !\left\langle nz\left\vert \mathcal{E}%
_{f}^{-1}\hat{B}\mathcal{E}_{f}^{-1}\right\vert nz\right\rangle  \notag \\
&=:&\mathrm{e}^{\left( 1+s\right) \left\vert z\right\vert ^{2}}N_{z,f}^{2}%
\left[ f^{2}\left( n\right) \right] !K_{B\left( f\right) }\left( nz\right) .
\notag
\end{eqnarray}
where 
\begin{equation}
\hat{B}\left( f\right) :=\mathcal{E}_{f}^{-1}\hat{B}\mathcal{E}_{f}^{-1}.
\end{equation}
Again, the $\left( f,s\right) -$deformed $K-$function of an operator $\hat{B}
$ is the $K-$function of the $f-$deformed operator $\hat{B}\left( f\right) $
times a known $c-$number. Then, the reconstruction formula for any bounded
operator $\hat{B}$\ in the $\left( f,s\right) -$deformed photon number
tomography reads: 
\begin{equation}
\hat{B}=\sum\limits_{n=0}^{\infty }\int \frac{d^{2}z}{\pi }\mathcal{E}_{f}%
\hat{G}_{ \lambda }(nz)\mathcal{E}_{f}K_{B\left( f\right)
}(nz)=:\sum\limits_{n=0}^{\infty }\int \frac{d^{2}z}{\pi }\hat{G}_{\lambda
}^{f,s}(nz)K_{B}^{f,s}(nz).
\end{equation}
with 
\begin{eqnarray}
\hat{G}_{\lambda }^{f,s}(nz) &=&\frac{\mathrm{e}^{-\left( 1+s\right)
\left\vert z\right\vert ^{2}}}{\left[ f^{2}\left( n\right) \right]
!N_{z,f}^{2}}\mathcal{E}_{f}\hat{G}_{ \lambda }(nz)\mathcal{E}_{f}  \notag \\
&=&\frac{\mathrm{e}^{-\left( 1+s\right) \left\vert z\right\vert ^{2}}}{\left[
f^{2}\left( n\right) \right] !N_{z,f}^{2}}\frac{4}{1-\lambda ^{2}}\left( 
\frac{\lambda +1}{\lambda -1}\right) ^{n}\mathcal{E}_{f}^{2}\mathcal{D}%
_{f}\left( z\right) \left( \frac{\lambda -1}{\lambda +1}\right) ^{\hat{A}%
_{f}^{\dag }\hat{A}}\mathcal{D}_{f}^{-1}\left( z\right) ,  \notag
\end{eqnarray}
which is readily obtained by using 
\begin{equation}
\mathcal{E}_{f}\mathcal{D}\left( z\right) \hat{a}^{\dagger }\hat{a}\mathcal{D%
}^{\dagger }\left( z\right) \mathcal{E}_{f}=\mathcal{E}_{f}^{2}\mathcal{D}%
_{f}\left( z\right) \hat{a}^{\dagger }\hat{a}\mathcal{D}_{f}\left( -z\right)
=\mathcal{E}_{f}^{2}\mathcal{D}_{f}\left( z\right) \hat{A}_{f}^{\dag }\hat{A}%
\mathcal{D}_{f}^{-1}\left( z\right) .
\end{equation}

\section{Conclusions}

To conclude, we recapitulate the new results of this work. We have
constructed for infinite dimensional Hilbert spaces the tomographic sets
based on coherent states and Fock states, respectively, and proved
decomposition of the identity operator acting on $B(\mathcal{H}),$ the space
of bounded operators on the underlying Hilbert space $\mathcal{H},$\ in
terms of rank-one projectors of the tomographic set and associated
Gram-Schmidt operators.

The properties of linear and non-linear $f-$oscillators were used to develop
new schemes of the coherent state tomography. The $f-$deformed tomographic
projectors were shown to yield a resolution of the (super-) unity simply by
means of a linear non-canonical operator which may be used to deform the
scalar product definition of $\mathcal{H}.$ The particular case of the
deformed oscillators, namely $q-$deformed oscillators, were studied as
example of the coherent state tomography.

We addressed the general problem of mapping an abstract Hilbert space on the
set of fair probability distributions (tomograms) describing quantum states
in the framework of probability picture of quantum mechanics.

We hope to extend the obtained results to the case of multimode quantum
systems and entangled states in future papers.

\noindent \textbf{Acknowledgements} V. I. Man'ko thanks for hospitality and
support INFN, Sezione di Napoli and University \textquotedblleft Federico
II\textquotedblright\ of Naples.\newline
E. C. G. Sudarshan's work was supported by U.S.Navy-Office of Naval
Research, Grants No. N00014-04-1-0336 and No. N00014-03-1-0639.

\section{Appendix}

\subsection{Checking the resolution of the unity for the coherent state
tomography}

In the position representation the reconstruction formula Eq. (\ref
{Areconstr}) reads: 
\begin{eqnarray}
\left\langle x\left\vert \hat{A}\right\vert x^{\prime }\right\rangle &=&\int 
\frac{d^{2}z^{\prime}}{\pi }\left\langle x\left\vert \hat{G}%
(z^{\prime})\right\vert x^{\prime }\right\rangle K_{A}(z^{\prime}) \\
&=&\int dydy^{\prime }\left[ \int \frac{d^{2}z^{\prime}}{\pi }\left\langle
x\left\vert \hat{G}(z^{\prime})\right\vert x^{\prime }\right\rangle
\left\langle z^{\prime}|y\right\rangle \left\langle y^{\prime
}|z^{\prime}\right\rangle \right] \left\langle y\left\vert \hat{A}%
\right\vert y^{\prime }\right\rangle .  \notag
\end{eqnarray}
So, we check the resolution of the unity generated by the CS tomographic set
by showing that 
\begin{equation}
\int \frac{d^{2}z^{\prime}}{\pi }\left\langle x\left\vert \hat{G}%
(z^{\prime})\right\vert x^{\prime }\right\rangle \left\langle y^{\prime
}|z^{\prime}\right\rangle \left\langle z^{\prime}|y\right\rangle =\delta
(x-y)\delta (x^{\prime }-y^{\prime }).  \label{checkCS}
\end{equation}

Recalling that 
\begin{eqnarray}
&&\left\langle y^{\prime }\left| z^{\prime }\right\rangle \left\langle
z^{\prime }\right| y\right\rangle  \notag \\
=\frac{1}{\sqrt{\pi }} &&\exp \left[ i\sqrt{2}z_{I}^{\prime }(y^{\prime }-y)%
\right] \exp \left( -\frac{1}{2}\left[ (y-\sqrt{2}z_{R}^{\prime
})^{2}+(y^{\prime }-\sqrt{2}z_{R}^{\prime })^{2}\right] \right)
\end{eqnarray}
and 
\begin{equation}
\left\langle x\left| \hat{G}(z^{\prime })\right| x^{\prime }\right\rangle
=\int \frac{d^{2}z}{2\pi }\int \frac{d\xi d\eta }{2\pi }\mathrm{e}^{(\xi
^{2}+\eta ^{2})/4}\mathrm{e}^{i\left[ \xi \left( z_{R}-z_{R}^{\prime
}\right) +\eta \left( z_{I}-z_{I}^{\prime }\right) \right] }\left\langle
x\left| z\right\rangle \left\langle z\right| x^{\prime }\right\rangle
\end{equation}
we may evaluate first the integral over $z_{I}$ 
\begin{equation}
\int \frac{dz_{I}}{2\pi }\exp \left[ iz_{I}(\sqrt{2}\left( x-x^{\prime
}\right) +\eta )\right] =\delta (\sqrt{2}\left( x-x^{\prime }\right) +\eta )
\end{equation}
and the integral over $z_{I}^{\prime }$%
\begin{equation}
\int \frac{dz_{I}^{\prime }}{2\pi }\exp \left[ iz_{I}^{\prime }(\sqrt{2}%
\left( y^{\prime }-y\right) -\eta )\right] =\delta (\sqrt{2}\left( y^{\prime
}-y\right) -\eta )\quad .
\end{equation}
Then, apart from the pre-factor 
\begin{equation}
\frac{2}{\pi }\exp \left( -\frac{1}{2}\left( \left[ y^{2}+y^{\prime 2}\right]
+\left[ x^{2}+x^{\prime 2}\right] \right) \right) ,
\end{equation}
we have to evaluate the integral 
\begin{eqnarray}
&&\int dz_{R}dz_{R}^{\prime }\int \frac{d\xi d\eta }{2\pi }\mathrm{e}^{(\xi
^{2}+\eta ^{2})/4}\mathrm{e}^{i\left[ \xi \left( z_{R}-z_{R}^{\prime
}\right) \right] }\delta (\sqrt{2}\left( x-x^{\prime }\right) +\eta )\delta (%
\sqrt{2}\left( y^{\prime }-y\right) -\eta )\times  \notag \\
&&\mathrm{e}^{\left( -2z_{R}^{\prime 2}+\sqrt{2}z_{R}^{\prime }(y+y^{\prime
})\right) }\mathrm{e}^{\left( -2z_{R}^{2}+\sqrt{2}z_{R}(x+x^{\prime
})\right) }\,\ .
\end{eqnarray}
The integral over $\eta $ yields 
\begin{eqnarray}
&&\int d\eta \mathrm{e}^{\eta ^{2}/4}\delta (\sqrt{2}\left( x-x^{\prime
}\right) +\eta )\delta (\sqrt{2}\left( y^{\prime }-y\right) -\eta ) \\
&=&\mathrm{e}^{\left( \frac{1}{4}\left[ y^{\prime }-y\right] ^{2}+\frac{1}{4}%
\left[ x^{\prime }-x\right] ^{2}\right) }\int d\eta \delta (\sqrt{2}\left(
x-x^{\prime }\right) +\eta )\delta (\sqrt{2}\left( y^{\prime }-y\right)
-\eta )  \notag \\
&=&\exp \left( \frac{1}{4}\left[ y^{\prime }-y\right] ^{2}+\frac{1}{4}\left[
x^{\prime }-x\right] ^{2}\right) \frac{1}{\sqrt{2}}\delta (x-x^{\prime
}+y^{\prime }-y),  \notag
\end{eqnarray}
so that the pre-factor becomes 
\begin{equation}
\frac{2}{\pi }\exp \left( -\frac{1}{4}\left( \left[ y^{\prime }+y\right]
^{2}+\left[ x^{\prime }+x\right] ^{2}\right) \right) \frac{1}{\sqrt{2}}%
\delta (x-x^{\prime }+y^{\prime }-y),
\end{equation}
while the gaussian integrals over $z_{R}$ and $z_{R}^{\prime }$ give 
\begin{eqnarray}
&&\int dz_{R}dz_{R}^{\prime }\mathrm{e}^{\left( -2z_{R}^{\prime
2}+z_{R}^{\prime }\left[ \sqrt{2}\left( y+y^{\prime }\right) -i\xi \right]
\right) }\mathrm{e}^{\left( -2z_{R}^{2}+z_{R}\left[ \sqrt{2}\left(
x+x^{\prime }\right) +i\xi \right] \right) } \\
&=&\frac{\pi }{2}\mathrm{e}^{\left[ \sqrt{2}\left( y+y^{\prime }\right)
-i\xi \right] ^{2}/8}\mathrm{e}^{\left[ \sqrt{2}\left( x+x^{\prime }\right)
+i\xi \right] ^{2}/8}  \notag \\
&=&\frac{\pi }{2}\exp \left( \frac{1}{4}\left( \left[ y+y^{\prime }\right]
^{2}+\left[ x+x^{\prime }\right] ^{2}\right) \right) \mathrm{e}^{-\xi ^{2}/4}%
\mathrm{e}^{i\xi \left( x+x^{\prime }-(y+y^{\prime })\right) \sqrt{2}/4}\,\ .
\notag
\end{eqnarray}
Collecting all the terms, the pre-factor becomes 
\begin{equation}
\frac{1}{\sqrt{2}}\delta (x-y-\left( x^{\prime }-y^{\prime }\right) )
\end{equation}
and the last integral to evaluate is 
\begin{equation}
\int \frac{d\xi }{2\pi }\mathrm{e}^{i\xi \left( x+x^{\prime }-(y+y^{\prime
})\right) \sqrt{2}/4}=2\sqrt{2}\delta \left( x+x^{\prime }-(y+y^{\prime
})\right) .
\end{equation}
So, we get the expected result 
\begin{equation}
2\delta (x-y-\left( x^{\prime }-y^{\prime }\right) )\delta \left(
x-y+(x^{\prime }-y^{\prime })\right) =\delta \left( x-y\right) \delta
(x^{\prime }-y^{\prime }).
\end{equation}

\subsection{Checking the resolution of the unity for the photon number
tomography}

Our aim is to check, for any allowed value of $\lambda ,$ the matrix form of
the resolution of unity for the photon number tomography, Eq.(\ref{Photoris}%
), which in the position representation reads 
\begin{equation}
\sum\limits_{n=0}^{\infty }\int \frac{d^{2}z}{\pi }\left\langle y^{\prime
}\left\vert nz\right\rangle \left\langle nz\right\vert x^{\prime
}\right\rangle \left\langle x|\hat{G}_{ \lambda }\left( n,z\right)
|y\right\rangle =\delta (x-x^{\prime })\delta (y-y^{\prime }).
\label{Photondecomp}
\end{equation}
With $z=\left( \nu +i\mu \right) /\sqrt{2},$ in the position representation $%
\left\langle y|nz\right\rangle $ is 
\begin{eqnarray}
\int dq\left\langle y|\mathcal{D}\left( z\right) |q\right\rangle
\left\langle q|n\right\rangle &=&\int dq\delta \left( y-q-\nu \right) \exp %
\left[ i\left( \mu q+\mu \nu /2\right) \right] \left\langle q|n\right\rangle
\notag \\
&=&\exp \left[ i\left( \mu y-\mu \nu /2\right) \right] \left\langle y-\nu
|n\right\rangle ,  \label{alfaenne}
\end{eqnarray}
where the $n$-th Hermite function $\left\langle q|n\right\rangle $ is 
\begin{equation}
\left\langle q|n\right\rangle =(\sqrt{\pi }2^{n}n!)^{-1/2}\exp (-\frac{1}{2}%
q^{2})H_{n}(q).  \label{accaenne}
\end{equation}
As we have shown in Ref.\cite{PLA2}, the matrix element of the Gram-Schmidt
operator $\hat{G}_{ \lambda }$ eventually reads: 
\begin{eqnarray}
\left\langle x|\hat{G}_{ \lambda }\left( n,z\right) |y\right\rangle &=&4\sin
^{2}\left( \frac{\tau }{2}\right) \frac{\mathrm{e}^{i\tau (n+1/2)}}{\sqrt{%
2\pi i\sin \tau }}\exp \left[ i\mu \left( y-x\right) \right] \times
\label{matrphonk} \\
\exp &&\left( i\left[ \left( \left( x-\nu \right) ^{2}+\left( y-\nu \right)
^{2}\right) \frac{\cos \tau }{2\sin \tau }-\frac{\left( x-\nu \right) \left(
y-\nu \right) }{\sin \tau }\right] \right)  \notag
\end{eqnarray}
where we have put 
\begin{equation*}
\frac{\lambda +1}{\lambda -1}=\mathrm{e}^{i\tau },\left( \tau |_{\lambda
=0}=\pi \right) \Rightarrow \ \lambda =\frac{\mathrm{e}^{i\tau }+1}{\mathrm{e%
}^{i\tau }-1}=-i\cot \frac{\tau }{2};\ \frac{4}{1-\lambda ^{2}}=4\sin ^{2}%
\frac{\tau }{2}.
\end{equation*}
Formulae (\ref{Phonumk}), (\ref{matrphonk}) correct the corresponding
expressions given in Ref. \cite{PLA2}.

Bearing in mind the expression of $\left\langle y^{\prime }|nz\right\rangle $
given by Eq.s(\ref{alfaenne},\ref{accaenne}) and recalling that $z=\left(
\nu +i\mu \right) /\sqrt{2},$ so that $d^{2}z=d\mu d\nu /2,$ we have to
evaluate 
\begin{equation}
\sum\limits_{n=0}^{\infty }\int \frac{d\mu d\nu }{2\pi }\exp \left[ i\mu
(y^{\prime }-x^{\prime })\right] \left\langle y^{\prime }-\nu \left\vert
n\right\rangle \left\langle n\right\vert x^{\prime }-\nu \right\rangle
\left\langle x|G^{\left( \lambda \right) }\left( n,z\right) |y\right\rangle .
\end{equation}
Extracting the factor $\exp \left[ i\mu (y-x)\right] $ from $\left\langle x|%
\hat{G}_{ \lambda }\left( n,z\right) |y\right\rangle $, integration over $%
\mu $ yields 
\begin{equation}
\int \frac{d\mu }{2\pi }\exp \left[ i\mu (x^{\prime }-y^{\prime })\right]
\exp \left[ i\mu (y-x)\right] =\delta (x^{\prime }-x-(y^{\prime }-y)).
\end{equation}

It is possible to sum the series by means of the Mehler's formula, which is
derived in the next subsection: 
\begin{eqnarray}
\sum\limits_{n=0}^{\infty }\left( \frac{z}{2}\right) ^{n}\frac{%
H_{n}(x)H_{n}(y)}{n!} &=&\frac{1}{\sqrt{1-z^{2}}}\exp \left[ \frac{%
z^{2}(x^{2}+y^{2})-2zxy}{z^{2}-1}\right] ,  \notag \\
(\left\vert z\right\vert &\leq &1,\quad z\neq \pm 1).
\end{eqnarray}
In our case $z=\mathrm{e}^{i\tau },$ so 
\begin{equation*}
\frac{z}{z^{2}-1}=\frac{1}{z-z^{-1}}=\frac{-i}{2\sin \tau }\quad ;\quad 
\frac{1}{\sqrt{1-z^{2}}}=\frac{z^{-1/2}}{\sqrt{z^{-1}-z}}=\frac{\mathrm{e}%
^{-i\tau /2}}{\sqrt{-2i\sin \tau }}
\end{equation*}
and the Mehler's formula reads 
\begin{equation}
\frac{\mathrm{e}^{-i\tau /2}}{\sqrt{-2i\sin \tau }}\exp \left[ \frac{-i}{%
2\sin \tau }\left\{ \mathrm{e}^{i\tau }\left[ (x^{\prime }-\nu
)^{2}+(y^{\prime }-\nu )^{2}\right] -2(x^{\prime }-\nu )(y^{\prime }-\nu
)\right\} \right] .
\end{equation}
We are left with a prefactor 
\begin{equation}
4\sin ^{2}\left( \frac{\tau }{2}\right) \frac{1}{\sqrt{\pi }}\frac{\mathrm{e}%
^{-i\tau /2}}{\sqrt{-2i\sin \tau }}\frac{\mathrm{e}^{i\tau /2}}{\sqrt{2\pi
i\sin \tau }}=\frac{1}{\pi }\left\vert \tan \frac{\tau }{2}\right\vert
\end{equation}
and the integral 
\begin{eqnarray}
&&\int d\nu \exp \left( -\frac{\sin \tau +i\mathrm{e}^{i\tau }}{2\sin \tau }%
\left[ \left( x^{\prime }-\nu \right) ^{2}+\left( y^{\prime }-\nu \right)
^{2}\right] +i\frac{(x^{\prime }-\nu )(y^{\prime }-\nu )}{\sin \tau }\right)
\times  \notag \\
&&\exp \left( i\frac{\cos \tau }{2\sin \tau }\left[ \left( x-\nu \right)
^{2}+\left( y-\nu \right) ^{2}\right] -i\frac{\left( x-\nu \right) \left(
y-\nu \right) }{\sin \tau }\right) .
\end{eqnarray}
The coefficient of $\nu ^{2}$ vanishes: 
\begin{equation}
-\frac{\sin \tau +i\mathrm{e}^{i\tau }}{\sin \tau }+i\frac{\cos \tau }{\sin
\tau }=0.
\end{equation}
Then, apart from the prefactor 
\begin{equation}
\frac{\left\vert \tan \frac{\tau }{2}\right\vert }{\pi }\exp \left( -\frac{%
i\cos \tau }{2\sin \tau }\left[ x^{\prime 2}+y^{\prime 2}\right] +\frac{%
i\cos \tau }{2\sin \tau }\left[ x^{2}+y^{2}\right] +\frac{i(x^{\prime
}y^{\prime }-xy)}{\sin \tau }\right) ,
\end{equation}
we have 
\begin{eqnarray*}
&&\int d\nu \exp \left( \nu \left[ \frac{i\cos \tau }{\sin \tau }(x^{\prime
}+y^{\prime })-\frac{i\cos \tau }{\sin \tau }(x+y)-\frac{i(x^{\prime
}+y^{\prime }-(x+y))}{\sin \tau }\right] \right) \\
&=&\int d\nu \exp \left[ -i\nu (x^{\prime }+y^{\prime }-(x+y))\tan \frac{%
\tau }{2}\right] =\frac{2\pi }{\left\vert \tan \frac{\tau }{2}\right\vert }%
\delta (x^{\prime }+y^{\prime }-(x+y)).
\end{eqnarray*}
Hence we get the expected product of delta functions 
\begin{equation}
2\delta (x^{\prime }-x-(y^{\prime }-y))\delta (x^{\prime }+y^{\prime
}-(x+y))=\delta (x^{\prime }-x)\delta (y^{\prime }-y).
\end{equation}
In view of this result, we may assume $x^{\prime }=x,y^{\prime }=y$ in the
prefactor 
\begin{equation}
\exp \left( -\frac{i\cos \tau }{2\sin \tau }\left[ x^{\prime 2}+y^{\prime 2}%
\right] +\frac{i\cos \tau }{2\sin \tau }\left[ x^{2}+y^{2}\right] +\frac{%
i(x^{\prime }y^{\prime }-xy)}{\sin \tau }\right) \rightarrow 1,
\end{equation}
so that we obtain just the resolution of unity of Eq.(\ref{Photondecomp}).

\subsection{A deduction of the Mehler's formula in the holomorphic
representation}

Here we give a deduction of the Mehler's formula (1866) by using the
holomorphic, i.e. Bargmann-Fock, representation of the harmonic oscillator
wave functions: 
\begin{eqnarray}
\left\langle z|\psi _{n}\right\rangle &=&\psi _{n}\left( z^{\ast }\right) =%
\frac{z^{\ast n}}{\sqrt{n!}},\quad z=z_{R}+iz_{I}; \\
\left\langle \psi _{m}|\psi _{n}\right\rangle &=&\int \frac{z^{m}}{\sqrt{m!}}%
\frac{z^{\ast n}}{\sqrt{n!}}d\mu _{a}=\delta _{nm},\quad d\mu _{a}:=\mathrm{e%
}^{-z^{\ast }z}\frac{dz_{R}dz_{I}}{\pi }=\mathrm{e}^{-z^{\ast }z}\frac{%
dz^{\ast }dz}{2\pi i}.  \notag
\end{eqnarray}
Representation of Dirac's delta function reads: 
\begin{equation}
f(z^{\ast })=\int \mathrm{e}^{z^{\ast }z^{\prime }}f(z^{\prime \ast })d\mu
_{z^{\prime }}\,\ .
\end{equation}
We use this representation of the delta function for evaluating the integral 
\begin{eqnarray}
J(x,y;\zeta ) &=&\int \mathrm{e}^{\sqrt{2}xz-z^{2}/2}\mathrm{e}^{\zeta
z^{\ast }z^{\prime }}\mathrm{e}^{\sqrt{2}yz^{\prime \ast }-z^{\prime \ast
2}/2}d\mu _{z}d\mu _{z^{\prime }}  \notag \\
&=&\int \mathrm{e}^{\sqrt{2}xz-z^{2}/2}\mathrm{e}^{\sqrt{2}\zeta yz^{\ast
}-\zeta ^{2}z^{\ast 2}/2}d\mu _{z}\,\ .  \label{J}
\end{eqnarray}
Then we get for $J(x,y;\zeta )$ the gaussian integral: 
\begin{eqnarray}
&&\int \exp \left[ -\frac{1}{2}\left( z,z^{\ast }\right) \left[ 
\begin{array}{cc}
1 & 1 \\ 
1 & \zeta ^{2}
\end{array}
\right] \left( 
\begin{array}{c}
z \\ 
z^{\ast }
\end{array}
\right) +\sqrt{2}\left( x,\zeta y\right) \left( 
\begin{array}{c}
z \\ 
z^{\ast }
\end{array}
\right) \right] \frac{dz^{\ast }dz}{2\pi i}  \notag \\
&=&\left[ -\det \left[ 
\begin{array}{cc}
1 & 1 \\ 
1 & \zeta ^{2}
\end{array}
\right] \right] ^{-\frac{1}{2}}\exp \left[ \left( x,\zeta y\right) \left[ 
\begin{array}{cc}
1 & 1 \\ 
1 & \zeta ^{2}
\end{array}
\right] ^{-1}\left( 
\begin{array}{c}
x \\ 
\zeta y
\end{array}
\right) \right] .
\end{eqnarray}
Eventually 
\begin{equation}
J(x,y;\zeta )=\frac{1}{\sqrt{1-\zeta ^{2}}}\exp \left[ \frac{\zeta
^{2}\left( x^{2}+y^{2}\right) -2\zeta xy}{\zeta ^{2}-1}\right] .
\label{Mehler1}
\end{equation}
On the other side, remembering that the generating function of the Hermite
polynomials in the position representation is 
\begin{equation}
\mathrm{e}^{\sqrt{2}xz-z^{2}/2}=\sum_{n=0}^{\infty }\frac{H_{n}\left(
x\right) }{n!}\left( \frac{z}{\sqrt{2}}\right) ^{n},
\end{equation}
we have that Eq. (\ref{J}) also reads: 
\begin{equation}
J(x,y;\zeta )=\sum_{n,m=0}^{\infty }\frac{H_{n}\left( x\right) H_{m}\left(
y\right) }{2^{\left( n+m\right) /2}n!m!}\int z^{n}\mathrm{e}^{\zeta z^{\ast
}z^{\prime }}\left( z^{\prime \ast }\right) ^{m}d\mu _{z}d\mu _{z^{\prime
}}\,\ .
\end{equation}
Again we use the Dirac's delta and get for $J(x,y;\zeta ):$%
\begin{equation}
\sum_{n,m=0}^{\infty }\frac{H_{n}\left( x\right) H_{m}\left( y\right) }{%
2^{\left( n+m\right) /2}n!m!}\int z^{n}\left( \zeta z^{\ast }\right)
^{m}d\mu _{z}=\sum_{n,m=0}^{\infty }\frac{\zeta ^{m}H_{n}\left( x\right)
H_{m}\left( y\right) }{2^{\left( n+m\right) /2}\sqrt{n!m!}}\left\langle \psi
_{n}|\psi _{m}\right\rangle
\end{equation}
By comparing this expression of $J(x,y;\zeta )$ with the previous one of Eq.
(\ref{Mehler1}) we finally obtain the Mehler's formula 
\begin{equation}
\sum_{n=0}^{\infty }\left( \frac{\zeta }{2}\right) ^{n}\frac{H_{n}\left(
x\right) H_{n}\left( y\right) }{n!}=\frac{1}{\sqrt{1-\zeta ^{2}}}\exp \left[ 
\frac{\zeta ^{2}\left( x^{2}+y^{2}\right) -2\zeta xy}{\zeta ^{2}-1}\right] .
\end{equation}
The radius of convergence of the series is $1$, but from the previous
evaluation we get that the values $\left| \zeta \right| =1,\zeta \neq \pm 1$
are allowed because 
\begin{equation}
\det \left[ 
\begin{array}{cc}
1 & 1 \\ 
1 & \zeta ^{2}
\end{array}
\right] \neq 0.
\end{equation}
The limits $\zeta \rightarrow \pm 1$ are 
\begin{equation}
\lim_{\zeta \rightarrow \pm 1}\frac{1}{\sqrt{1-\zeta ^{2}}}\exp \left[ \frac{%
\zeta ^{2}\left( x^{2}+y^{2}\right) -2\zeta xy}{\zeta ^{2}-1}\right] =\sqrt{%
\pi }\delta \left( x\mp y\right) .
\end{equation}

\end{document}